# Intelligent Bees for QoS Routing in Networks-on-Chip

Peibo Xie
*State Key Laboratory of Integrated Service Networks, Xidian University*
Xi'an, China
peiboxie@gmail.com

Huaxi Gu
*State Key Laboratory of Integrated Service Networks, Xidian University*
Xi'an, China
hxgu@xidian.edu.cn

*Abstract* - Networks-on-Chip (NoCs) for future many-core processor platforms integrate more and more heterogeneous components of different types and many real-time and latency-sensitive applications can run on a single chip concurrently. The reconfigurable FPGA and reconfigurable NoCs have emerged for the purpose of reusability. Those types' traffics within NoCs exhibit diverse, burst, and unpredictable communication patterns. QoS guaranteed mechanisms are necessary to provide guaranteed throughput (GT) or guaranteed bandwidth (GB) performance for NoCs. In this paper, we propose a QoS routing algorithm inspired by bees' foraging behaviors to provide guaranteed bandwidth performance. Virtual circuits and Spatial Division Multiplexing are employed to maintain available paths for different type's traffics.

*Index Terms - NoCs; QoS Routing; Virtual Circuit; SDM.*

## I. INTRODUCTION

Networks-on-Chip introduce a new design methodology of interconnection network for System-on-Chip design. With the increasing complexity of NoCs [1-3], more and more heterogeneous components of different types are implemented. Large number of multimedia applications can run on a single chip concurrently. Different applications have different QoS con-strains, and more and more applications are becoming sophisticated for bandwidth, end-to-end delay, and security and other service parameters. Meanwhile, reconfigurable FPGA platforms[4,5] and reconfigurable NoCs [6,7] have emerged. They allow different applications to be configured on the same chip without adding extra modules and even a single component can perform various functions such as different Turbo Decoding standards for a 3G phone. Traffics within NoCs implementing multimedia applications exhibit diverse, burst, and unpredictable communication patterns. Traditional best-effort service model, which treats all the traffics fairly, can no longer provide timely and effective service for those traffics which has lower bound of bandwidth or upper bound of delay. It results in the overall network performance degradation. Therefore, new quality of service guaranteed mechanisms are pressingly needed to provide guaranteed throughput (GT) or guaranteed bandwidth (GB) performance for NoC.

There have been much research works concerning on how to guarantee QoS of different traffics. They can be divided into three mechanisms according to the means of realization: resource reservation, priority assignment and scheduling, and QoS routing. Resource reservation needs to reserve required network resources such as link bandwidths and buffers in advance before sending the data to establish a complete connection. Resource reservation can provide better quality of service which is known as hard real-time guarantees. The most simple one is circuit switching[8,9]. In addition to circuit switching, Time Division Multiplexing (TDM) is another popular research topic. Æthereal[10] and Nostrum[11] are typical resource reservation examples which employ TDM virtual circuits to establish connections for communication node pairs. But TDM virtual circuit requires that the overall network use a unified clock frequency and full synchronization of all components, which in the actual design and applications are difficult to achieve. Meanwhile, resource reservation, however, is a worst-case model for modeling the traffic which may lead to low resource utilization.

Priority assignment and scheduling [12-14] is another mechanism that is often employed by multi-hop packet switching network. It decides and classifies different traffics into different levels according to a certain classification rules. Packets of different levels compete for the arbitration, and the packets with high priorities have the right to utilize the ports and links beyond the packets with low priorities. Priority assignment and scheduling is often regarded as a type of soft real-time guarantees because it allows some traffic in the network to exceed the limits of its maximum allowed delay with a slight performance degradation. So, it gives bad QoS guarantees, but higher utilization of network resources compared with resource reservation.

As mentioned above, most research works on QoS within NoC are concerned on how to reverse network resources efficiently or assign traffics' priorities properly to achieve better performance. However, few people are involved in designing efficient routing algorithms to achieve the objective of guaranteeing QoS. How to pick satisfied paths for the accessed traffics to meet the QoS constraints, while ensuring efficient utilization of network resources, is a novel research aspect. QoS Routing guarantees QoS of traffics in the form of routing algorithm. Apart from the correct transmission of datas, the QoS routing algorithm is also responsible for guaranteeing to meet the QoS constraints of traffics. It requires QoS routing algorithm can intelligently identify the urgency of the traffics and treat them separately. In this paper we propose a new QoS routing algorithm inspired by bees' foraging behaviors.

This paper is organized as follows. In Section 2, we introduce the bees' swarm intelligence in nature and summarize its implementations in solving problems such as TSP, task scheduling and network routing algorithms. After that, in

Section 3, we present our beehive-based QoS routing algorithm. Finally, we conclude the paper in Section 4.

## II. BEES IN NATURE

Bees are typical social insects as well as ants. A queen bee, numerous drone bees and worker bees compose the classical bee colony. Different members in the hive have different routine tasks to maintain the colony's prosperity. Worker bees, responsible for nesting, collecting nectar and feeding larvaes and so on, cover almost the whole colony. Among them, the bees, which fly out to search for and collect nectar sources, are the most interesting ones.

Every day, there are thousands of scout bees flying away from their hive to search for nectar. Each scout goes its own way without communicating with others during their searching process. On finding a nectar source, it flies back to the hive at once informing other scouts and the following bees the nectar information such as direction, distance and quality of the nectar by the means of dancing. Different distance leads to different dancing styles. When the nectar is close to the hive (<50 meters), the scout will play round dancing, or else it will perform the waggle dancing in the case of more than 50 meters. Longer duration of dancing also means better quality of nectar. Meanwhile, the other bees watch the dancing and determine which bee they will follow. To avoid the information explosion induced by much dancing, each following bee is allowed to watch at most three scouts' dancing. As for the scout bees back to the hive, not all of them will play dancing in the dancing floor due to their nectar qualities. Some of them will also watch other scout bees' dancing and discard the nectar found by it to be a following bee while some of them will fly out again to search for new nectar without telling others about the nectar information. The left will play dancing and recruit a number of following bees to fetch the nectar.

Each bee's simple behavior constitutes the whole colony's complicated swarm intelligence. This information exchange method, in which bees communicate with each other by dancing in the dancing flooring, has no use of any central control mechanism. Scouts and following bees evaluate nectar's quality according to information gained from other scouts. It isn't necessary for each bee to be aware of the global information of environment. And the nectar, which has a better quality and shorter diatance, will be more favored. So, it's helpful to search for more nectar with good quality and increase the amount of nectar the whole bee colony can gain.

Some researchers have devoted themselves into discovering the mechanism existed in bee colony to solve some practical problems and have obtained some achievements. They have found it gained better performance than Ant Colony Optimization in solving the problem such as traveling salesman problem and complicated tasks scheduling and so on. Bee colony's no central control and unique evaluation method are needed by designing network routing algorithms. Recently, a group of researchers in Germany have utilized bees' swarm intelligence to solve routing problem in the fields of telecommunication[15], mobile ad-hoc[16,17] and wireless sensor networks[18]. But there is no attempt to treat the QoS routing problem in Network-on-Chip using bees' swarm intelligence.

## III. QOS ROUTING ALGORITHM INSPIRED BY BEES' SWARM INTELLIGENCE

### A. The Problem Formation

A typical NoC is often described as a graph composed by various IP blocks and links connecting them. In the graph $G(V,E)$, $V$ is node set and $E$ the edge set. Let $v_s \in V$ and $v_d \in \|V - \|s\|\|$ represent the traffic generator and sink respectively. $P = (v_1, v_2, ..., v_k)$ is called a path from $v_1$ to $v_k$ if there exists $v_i, v_{i+1} \in E$ ($i = 1, 2, ..., k-1$).

The target of QoS routing is to find out an available path from the traffic generator $v_s$ to the sink node $v_d$ that meets all the QoS constraints such as traffic bandwidth, upper bound of end-to-end delay, delay variance and so on. It is, however, not advisable and practical to consider the overall constraints while designing an efficient routing algorithm. Sometimes, existed constraints may conflict with each other, so that the proper path does not exist potentially. Even if there exists one, it would take too much time to calculate an appropriate path that meets all the constraints. And the hardware cost supporting the algorithm may be too large to adopt. Hence, we only consider several unignorable constraints to be satisfied in our routing algorithm, among which are traffic bandwidth and upper bound of end-to-end delay. While for the delay variance, Virtual circuit and SDM are applied to packet's ordered transmission provided that valid delay is guaranteed. Let $B_p$ demote the bandwidth constraint. The objective of routing algorithm with QoS constraints is to choose an optimal path from $v_s$ to $v_d$ that satisfy following conditions:

$$bandwidth(p(v_s, v_d)) \geq B_p \qquad (1)$$

$$\min(delay(p(v_s, v_d))) \qquad (2)$$

### B. QoS Routing Algorithm Inspired by Bees' Swarm Intelligence

Our proposed routing algorithm employs two gordian techniques: connection-oriented virtual circuit and Spatial Division Multiplexing (SDM)[19]. The virtual circuit is used to establish a connection between the source and destination nodes before message's transmission. The connection isn't a physical connection but a logical connection. Different from the circuit switching which occupies the overall bandwidth of links reserved by the connection, virtual circuit allow multiple traffics to share link band-widths in the form of TDM or SDM. SDM is based on the fact that Networks-on-Chip links are physically made of a set of wires. It consists of allocating only a sub-set of the link wires to a given virtual circuit and allows multiple traffics to share the link wires simultaneously.

In[17], the authors regard the traffic generators as beehives while the sink nodes are the nectar sources. Different from the definitions made by them, we consider the traffic generators as nectar sources and the sink nodes the beehives. In general, it demands two approaches to establish a connection completely when using a routing algorithm adopting the connection-oriented pattern. In our routing algorithm, the forward bees fly from the nectar to inform other bees the position and proper path to the nectars firstly. The earliest three forward bees will become backward bees to fly back to the nectar. During their flight, the backward bees reserve necessary network resources such as router buffer and link bandwidth. The backward bee that goes back to the source node earliest will be accepted by the source and the message will follow the path created by it.

When new traffic generates by nodes randomly, obtain the source and destination addresses and the traffic's QoS constraints; the proposed QoS routing algorithm based on bees' intelligence is described as follows: The source node broadcasts forward bees to all its neighbor nodes to find the available path to the destination node. Each forward bee has the following format:

| Source Address | Destination Address | Hop Counter | Required Bandwidth | Port List |
|---|---|---|---|---|

Fig. 1 The forward bee format.

The "Hop Counter" domain records the total hops that the bee packet has routed. And its initial value is set to be 0. We restrict each forward bee's maximum routing hops to avoid the livelock caused by the forward bee's unlimited random walks in the network. In our algorithm, its value is labeled as twice that of euclidean distance between the source and destination nodes. The bandwidth the traffic requires is recorded in the "Required Bandwidth" domain in order that the forward bees can decide which node they will choose to route. The "Port List" domain is used to register the nodes the forward bees have passed by, and it is empty at the initial state. In order to minimize the size of the forward bees when the scale of network expands, output ports that the bee selects are recorded instead of nodes' addresses.

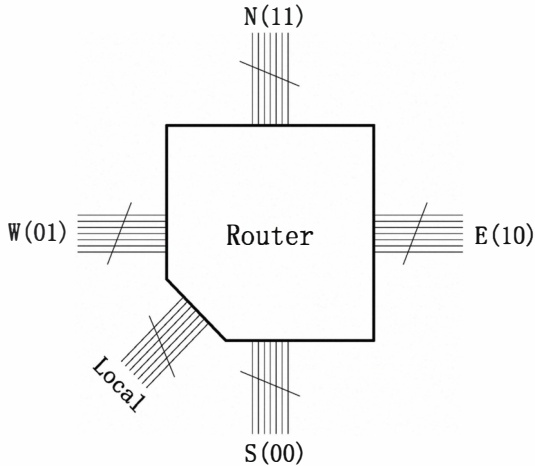

Fig. 2 Encoding mechanism of four port of a mesh router.

For Mesh topology, there exit four ports except the local port for a router. Binary combinations "00", "01", "11", "10" represent the south, west, north and east port of each node respectively in fig.2. When the bee selects a neighbor nodes which is in the east of the current node, the successor node is represented by binary combi-nation"10" in the forward bee's "Port List". There exist two proper paths from the source to the destination node in fig.3. The blue one is represented by the binary combination "10101010101000000000000000", and the yellow one is represented by "10101010000000101-00000000010". It only needs 28bits for the "Port List". As the network scales, the "Port List" domain is independent of the node's address length.

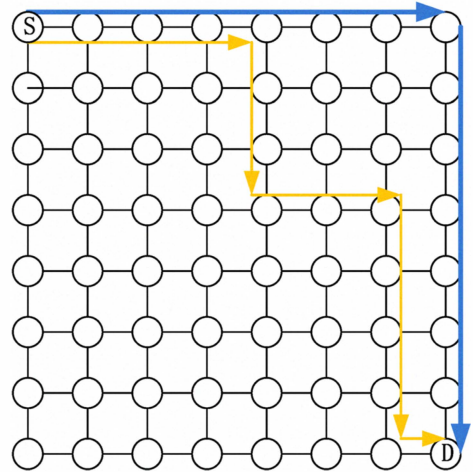

Fig. 3 Two different paths between S and D.

The routing algorithm is described in fig.4. The earliest three forward bees arriving at the destination node reverse their "Port Lists" and each bit in the port lists "xors" with binary number '1'. Other subsequent forward bees will be killed when they arrived. Then, three new packets called backward bees are generated to allocate the required bandwidth from the destination to source node. The backward bee that goes back to the source node earliest will be accepted by the source and the message will follow the path created by it. The remaining two bees will be discarded and the bandwidth they have allocated will be released. As soon as the message has been transmitted completely, the source node sends a unique small packet to tear down the connection between the source and destination, to release the occupied bandwidth, and to remove the corresponding flow in each node's flow table.

In the algorithm, since there exist a large number of forward bees belonging to the same message, it is a waste of network resources and affects other traffics' prompt trans-mission if each forward bee reserves the necessary link band-width during their flight. The forward bees therefore are only responsible for discovering the proper path back to the beehive but not reserving link bandwidth. The backward bees take the responsibilities to do it.

The condition of the whole network varies from time to time notably; and the link which meets the bandwidth constraint currently may become congested later. So the strategy that adopts three packets to allocate bandwidths simultaneously guarantees the success ratio for a connection. The failed backward bees will also release the link bandwidths they have reserved no matter whether they have arrived at the source node.

```
Inputs: coordinates of source node
        coordinates of destination node
        minimum link bandwidth needed by the traffic
Outputs: Selected path for the traffic
Procedure:
Source node broadcasts forward bees with uniform format;
the forward bees arrive at a node
if (the current node is the destination node)
    if (the forward bee is the earliest bees arrive at the
         destination node)
        the forward bee becomes a backward bee
    else
        kill the forward bee
    end if
else
    if (the forward bee had come before)
        kill the forward bee
    else
        if (hop counter has reached the maximum hops constraint)
            kill the forward bee
        else
            If (the incoming link bandwidth is more than or equal to
                 bandwidth_require)
                choose output ports and broadcast new forward bees
            else
                kill the forward bee
            end if
        end if
    end if
end if
```

Fig. 4 Pseudo code of our routing algorithm.

## IV. CONCLUSION

We proposed a QoS routing algorithm inspired by bees' swarm intelligence in this paper. The proposed routing algorithm simulates bees' foraging behaviors and unique information exchange pattern. And it is connection-oriented and employs virtual circuits and SDM techniques to allow multiple traffics to share links concurrently. The theoretical analysis suggests that our routing algorithm is suitable for traffics which last longer duration.

## ACKNOWLEDGMENT

This work was supported in part by the National Science Foundation of China under Grant No.60803038, 60725415 and 60971066, the special fund from State Key Lab (No.ISN090306), and the 111 Project under Grant No.B08038.

## REFERENCES

[1] Minje Jun, Sungroh Yoon and Eui-Young Chung, "Exploiting multiple switch libraries in topology synthesis of on-chip interconnection network," Proc. Design, Automation & Test in Europe Conference & Exhibition (DATE), 2010, Mar.2010, pp.1390-1395.

[2] Diemer Jonas, Ernst Rolf, "Back Suction: Service Guarantees for Latency-Sensitive On-chip Networks," Proc. 2010 Fourth ACM/IEEE International Symposium on Networks-on-Chip (NOCS), May.2010, pp. 155-162.

[3] Bo Niu, Osvaldo Simeone, Oren Somekh, Alexander M. Haimovich, "Ergodic and outage sum-rate of fading broadcast channels with 1-bit feedback," IEEE Trans. Veh. Technol., vol. 59, Issue 3, pp. 1282-1293, Mar. 2010.

[4] Byounghoon Kim, Youngmann Kim, Donggeon Lee, Sungwoo Tak, "A Reconfigurable NoC Platform Incorporating Real-Time Task Management Technique for H/W-S/W Codesign of Network Pro-tocols," Proc. International Symposium on Ubiquitous Multimedia Computing (UMC '08), Oct. 2008, pp. 238-243.

[5] Dario Cozzi, Claudia Fare, Alessandro Meroni, Vincenzo Rana, Marco Domenico Santambrogio, Donatella Sciuto, "Reconfigurable NoC design flow for multiple applications run-time mapping on FPGA devices," Proc. the 19th ACM Great Lakes symposium on VLSI, ACM Press, 2009, pp.421-424.

[6] Modarressi M, Sarbazi-Azad H, "Power-aware mapping for reconfigurable NoC architectures. in Computer Design," Proc. 25th International Conference on Computer Design (ICCD 2007), Oct. 2007, pp. 417-422.

[7] Stensgaard Mikkel Bystrup and Sparso Jens, "ReNoC: A Network-on-Chip Architecture with Reconfigurable Topology," Proc. the Second ACM/IEEE International Symposium on Networks-on-Chip(NoCS 2008), IEEE Computer Society, April. 2008, pp. 55-64.

[8] Wiklund D, Dake Liu, "SoCBUS: Switched Network on Chip for Hard Real Time Embedded Systems," Proc. Parallel and Distributed Processing Symposium, IEEE Computer Society, April. 2003, pp. 8-.

[9] Kuei-Chung Chang, Jih-Sheng Shen, Tien-Fu Chen, "Evaluation and design trade-offs between circuit-switched and packet-switched NOCs for application-specific SOCs," Proc. the 43rd annual Design Automation Conference, ACM Press, 2006, pp. 143-148.

[10] Goossens K, Dielissen J, Radulescu A, "AEthereal network on chip: concepts, architectures, and implementations," IEEE Design & Test of Computers, 2005. 22(5): pp. 414- 421, Sept.-Oct. 2005.

[11] Millberg M, Nilsson E, Thid R, Jantsch A, "Guaranteed bandwidth using looped containers in temporally disjoint networks within the nostrum network on chip," Proc. Design, Automation and Test in Europe Conference and Exhibition, Feb. 2004, vol 2, pp. 890-895.

[12] Evgeny Bolotin, Israel Cidon, Ran Ginosar, Avinoam Kolodny, "QNoC: QoS architecture and design process for network on chip," Journal of Systems Architecture, Elsevier North-Holland. Inc, vol.50, pp. 105–128, Feb. 2004.

[13] Horchani M, Atri M, Tourki R, "A SystemC QoS router design with virtual channels reservation in a wormhole-switched NoC," Proc. Third International Design and Test Workshop (IDT 2008), Dec. 2008, pp. 335-340.

[14] Zheng Shi, Burns A, "Priority Assignment for Real-Time Wormhole Communication in On-Chip Networks," Proc. Real-Time Systems Symposium, Dec. 2008, pp. 421-430.

[15] Horst F Wedde, Muddassar Farooq, "A comprehensive review of nature inspired routing algorithms for fixed telecommunication networks," Journal of System Architecture, Elsevier North-Holland. Inc, **52**(8): pp. 461-484, Aug. 2006.

[16] Horst F Wedde, Muddassar Farooq, "The wisdom of the hive applied to mobile ad-hoc networks," Proc. the IEEE Swarm Intelligence Symposium (SIS 2005), June 2005, pp. 341-348.

[17] Horst F Wedde, Muddassar Farooq, Thorsten Pannenbaecker, Bjoern Vogel, Christian Mueller, Johannes Meth, Rene Jeruschkat, "BeeAdHoc: an energy efficient routing algorithm for mobile ad-hoc networks inspired by bee behavior," Proc. the 2005 conference on Genetic and evolutionary computation (GECCO 2005), June 2005, pp. 153-160.

[18] Muhammad Saleem, Muddassar Farooq, "BeeSensor: A Bee-Inspired Power Aware Routing Protocol for Wireless Sensor Networks," Applications of Evolutionary Computing, Springer Berlin/Heidelberg, vol. 4448, pp. 81-90, June. 2007.

[19] Marchal P, Verkest D, Shickova A, Catthoor F, Robert F, Leroy A,"Spatial division multiplexing: a novel approach for guaranteed throughput on NoCs," Proc. Third IEEE/ACM/IFTP International Conference on Hardware/Software Coddling and System Synthesis (CODES+ISSS '05),Sept. 2005, pp. 81-86.